\documentclass[12pt,pra,aps,amssymb,amsfonts,amsmath,tightenlines]{revtex4}
\usepackage{dsfont}
\usepackage{graphicx}
\usepackage{amssymb,amsfonts,amsthm}
\usepackage{color}
\usepackage{verbatim}
\usepackage[normalem]{ulem}

\newtheorem{prop}{Proposition}
\newcommand{\half}{\mbox{$\textstyle \frac{1}{2}$}}

\newcommand{\re}{\mbox{$\rm e$}}

\newcommand{\rd}{\mbox{$\rm d$}}

\begin{document}

\title{General Theory of Geometric L\'evy Models \\ for Dynamic Asset Pricing}

\author{Dorje C.~Brody$^*$, Lane~P.~Hughston$^\dagger$, and Ewan~Mackie$^\dagger$}

\affiliation{$^*$Mathematical Sciences, Brunel University, Uxbridge UB8 3PH, UK \\ 
$^\dagger$Department of Mathematics, Imperial College London,
London SW7 2BZ, UK \\
$^\ddagger$Imperial College Business School, London SW7 2BZ, UK}

\date{\today}

\begin{abstract}
The geometric L\'evy model (GLM) is a natural generalisation of the geometric Brownian motion model (GBM) used in the derivation of the Black-Scholes formula. The theory of such models simplifies considerably if one takes a pricing kernel approach. In one dimension, once the underlying L\'evy process has been specified, the GLM has four parameters:  the initial price, the interest rate, the volatility, and the risk aversion. The pricing kernel is the product of a discount factor and a risk aversion martingale. For GBM, the risk aversion parameter is the market price of risk. For a GLM, this interpretation is not valid:  the excess rate of return is a nonlinear function of the volatility and the risk aversion. It is  shown that for positive volatility and risk aversion the excess rate of return above the interest rate is positive, and is increasing with respect to these variables. In the case of foreign exchange, Siegel's paradox implies that one can construct foreign exchange models for which the excess rate of return is positive both for the exchange rate and the inverse exchange rate. This condition is shown to hold for any geometric L\'evy model for foreign exchange in which volatility exceeds risk aversion.
\vspace{-0.2cm}
\begin{center}
\end{center}
\end{abstract}

\maketitle

\section{Investment-grade assets and excess rate of return}
\label{sec:I}

\noindent The goal of this paper is to understand better the nature of the 
``risk premium'' associated with jumps in asset prices. The idea is to work 
in a rather general setting, without being tied too much to any particular 
model. For mathematical simplicity, we shall assume that the dynamics of 
asset prices are driven by L\'evy processes. This already encompasses a 
large class of models---it includes, above all, all the Brownian motion based 
models---so we need not worry that we are being too restrictive. We aim to 
clarify the relation between risk, risk aversion, and the excess rate of return 
(above the interest rate) offered by risky assets in such a context. With this 
end in mind, let us recall the setup in the geometric Brownian motion (GBM) 
model. The GBM model is very simple, but it captures a number of the 
main features of the relation between risk, risk aversion, and the excess rate 
of return. We shall adopt a pricing kernel approach, 
which turns out to be particularly useful in the L\'evy case 
since it allows one to distinguish clearly between pricing issues and hedging 
issues.  See, for example, Cochrane (2005) for an overview of the 
application of pricing kernel methods.

In the one-dimensional case, we have a Brownian motion $\{W_t\}_{t\geq0}$ 
on a probability space $(\Omega,{\mathcal F},{\mathbb P})$, and the associated 
augmented filtration $\{{\mathcal F}_t\}_{t\geq0}$. Here ${\mathbb P}$ represents 
the  physical measure, and $\{{\mathcal F}_t\}$ is taken to be the market filtration. 
The model consists of: (a) a pricing kernel; and (b) a collection 
of one or more ``investment-grade'' assets. For simplicity, we assume for the 
time being that the assets under consideration pay no dividends over the time 
horizon considered. We relax this assumption in Section VI. 

The idea of an investment-grade asset is that it should offer a positive excess 
rate of return above the interest rate. There are respectable assets that do not 
have this property---such assets are typically held alongside investment-grade assets 
for hedging. One can check, for example, in the context of the Black-Scholes 
model, that the price process of a put option has a drift that is less than the 
interest rate.  The pricing kernel in the GBM model is
\begin{eqnarray}
\pi_t = \re^{-rt} \re^{-\lambda W_t - \frac{1}{2}\lambda^2t}, 
\label{eq:1}
\end{eqnarray}
where $r>0$ is the interest rate, and $\lambda >0$ is the risk aversion 
factor, both assumed to be constant. For a typical investment-grade asset we 
then have 
\begin{eqnarray}
S_t = S_0 \re^{(r+\lambda\sigma)t} \re^{\sigma W_t - \frac{1}{2}\sigma^2t},
\label{eq:2}
\end{eqnarray}
where $\sigma>0$ is the volatility. 
The term $\lambda\sigma$ is called the ``risk premium'' or ``excess rate of 
return'', and is clearly positive under the assumptions made. 

We observe that the risk premium is increasing with respect to both 
the volatility and the risk aversion. Since $\lambda\sigma$ is linear in 
each factor, we call $\lambda$ the ``market price of risk'' in the GBM model. It 
should be evident, however, that there is no \textit{a priori} reason why the 
excess rate of return should be bilinear. Indeed, we shall demonstrate that in 
a general L\'evy model 
the excess rate of return is a nonlinear function of $\lambda$ and $\sigma$. 
The reason that the pricing kernel is such a useful concept in finance is 
that market equilibrium and the absence of arbitrage are both built 
into the idea that the product of the pricing kernel with the price of any asset 
paying no dividend is a martingale.  In the GBM case, for example, we have
\begin{eqnarray}
\pi_tS_t = S_0 \re^{(\sigma-\lambda) W_t - \frac{1}{2}(\sigma-\lambda)^2t}. 
\end{eqnarray}
We shall use this property of the pricing kernel
to establish the general form of an arbitrage-free L\'evy-driven asset-pricing model. 
In Section~\ref{sec:2} we look at 
one-dimensional geometric L\'evy models with constant volatility and risk 
aversion, and in Section~\ref{sec:3} we establish the positivity and monotonicity of 
the risk premium for such models. In Section~\ref{sec:4} we consider models for 
foreign exchange, and establish conditions sufficient to ensure that both the 
exchange rate and the inverse exchange rate have a positive risk premium. 
Examples of geometric L\'evy models are studied in detail in Section~\ref{sec:5}, 
where we 
note the fact that, unlike the GBM case,  option prices in general depend on the 
risk aversion level; then in Section~\ref{sec:6} we extend the model to include 
dividends. In Section~\ref{sec:7} we consider models in which the 
market is driven by a vector of L\'evy processes, and in which the 
volatility and risk aversion are predictable processes. In both situations we 
establish  conditions sufficient to ensure the positivity of the excess rate of return. In 
conclusion we comment  on the advantage of the use of pricing kernel methods in 
the theory of L\'evy models, and how this allows us to unify earlier work on the subject, 
leading to a coherent framework for asset pricing.

\section{Pricing kernel approach to L\'evy models for asset prices}
\label{sec:2}

\noindent Let us construct a family of L\'evy models in the spirit of the GBM 
model. We shall call these geometric L\'evy models (GLMs). Here we consider 
the one-dimensional case. In what follows we shall assume that the reader is 
familiar with basic aspects of the theory of L\'evy processes, as represented 
for example in Appelbaum (2004), Bertoin (2004), Cont \& Tankov (2004), 
Kyprianou (2006), Protter (1990), Sato (1999), or Schoutens (2004). We recall 
that a L\'evy process on a probability space 
$(\Omega,{\mathcal F},{\mathbb P})$ is a process $\{X_t\}$ such that $X_0=0$, 
$X_t-X_s$ is independent of ${\mathcal F}_s$ for $t\geq s$ (independent 
increments), and 
\begin{eqnarray}
{\mathbb P}(X_t-X_s\leq y) = {\mathbb P}(X_{t+h}-X_{s+h}\leq y) 
\end{eqnarray}
(stationary increments). 
Here $\{{\mathcal F}_t\}$ denotes the augmented filtration generated by $\{X_t\}$. 
For $\{X_t\}$ to give rise to a GLM, we require that 
\begin{eqnarray}
{\mathbb E}[\re^{\alpha X_t}] < \infty  
\end{eqnarray}
for all $t\geq 0$, for  $\alpha$ in some  connected interval $A$ 
containing the origin.
Henceforth we consider L\'evy processes satisfying such a moment condition. 
It follows by the stationary and independent increments property that there exists 
a function $\psi(\alpha)$, the so-called L\'evy exponent, such that 
\begin{eqnarray}
{\mathbb E}[\re^{\alpha X_t}] = \re^{t\psi(\alpha)}  
\label{cumulant}
\end{eqnarray}
for $\alpha \in A$. The process $\{M_t\}$
defined by
\begin{eqnarray}
M_t = \re^{\alpha X_t-t\psi(\alpha)}   
\label {martingale}
\end{eqnarray}
is then called the 
geometric L\'evy martingale (or Esscher martingale) associated with $\{X_t\}$, with 
volatility $\alpha$.
Indeed, by the stationary and independent increments property 
we have:
\begin{eqnarray}
{\mathbb E}_s[M_t] = {\mathbb E}_s[\re^{\alpha X_t}]\, \re^{-t\psi(\alpha)} 
=  {\mathbb E}_s[\re^{\alpha (X_t-X_s)}] \, \re^{\alpha X_s-t\psi(\alpha)} 
= \re^{(t-s)\psi(\alpha)} \, \re^{\alpha X_s-t\psi(\alpha)} 
= M_s . 
\end{eqnarray}

Our geometric L\'evy model for asset prices will be put together as follows. First we 
construct the pricing kernel $\{\pi_t\}_{t\geq0}$. Let $\lambda>0$ and assume that 
$-\lambda\in A$. Then set 
\begin{eqnarray}
\pi_t = \re^{-rt} \re^{-\lambda X_t - t\psi(-\lambda)}.
\label{levy pricing kernel}
\end{eqnarray}
For a consistent pricing theory we require that the product of the pricing kernel and 
the asset price should be a martingale, which we shall assume is of the form
\begin{eqnarray}
\pi_tS_t = S_0\re^{\beta X_t - t\psi(\beta)}
\end{eqnarray}
for some $\beta\in A$. From the formulae above we deduce that
\begin{eqnarray}
S_t = S_0\, \re^{rt}\, \re^{(\beta+\lambda)X_t+t\psi(-\lambda)-t\psi(\beta)} 
= 
S_0 \, \re^{rt}\, \re^{\sigma X_t+t\psi(-\lambda)-t\psi(\sigma-\lambda)},
\label {price}
\end{eqnarray}
where $\sigma = \beta + \lambda$. We shall assume that  $\sigma > 0$ and that 
$\sigma\in A$. It follows that the asset price can be expressed 
in the form
\begin{eqnarray}
S_t = S_0 \, \re^{rt}\, \re^{R(\lambda,\sigma)t}\, \re^{\sigma X_t-t\psi(\sigma)},
\label{asset price}
\end{eqnarray}
where
\begin{eqnarray}
R(\lambda,\sigma) = \psi(\sigma) + \psi(-\lambda) - \psi(\sigma-\lambda). 
\label{R} 
\end{eqnarray}

\section{On the risk premium associated with L\'evy models}
\label{sec:3}

\noindent One sees that the function $R(\sigma, \lambda)$ is the risk premium,
that is to say, the excess rate of return above the interest rate. Indeed, 
we have:
\begin{eqnarray}
{\mathbb E}[S_t] = S_0 \, \re^{rt+R(\lambda,\sigma)t}.
\end{eqnarray}
The following result establishes a rather general property of geometric L\'evy 
models, and is indicative of why such models are of interest.

\begin{prop}\label{prop:1}
The excess rate of return in a geometric L\'evy model is positive, 
and is increasing with respect to the risk aversion and the 
volatility. 
\end{prop}

\noindent \textit{Proof}. 
We have
$\psi(\alpha) = t^{-1} \ln {\mathbb E}\left[ \re^{\alpha X_t}\right] $,
and thus
\begin{eqnarray}
\psi''(\alpha) 
=  \frac{1}{t} \frac{{\mathbb E}\left[(X_t-{\bar X}_t)^2\re^{\alpha X_t}\right]}
{{\mathbb E}\left[ \re^{\alpha X_t}\right]} , \quad {\rm where} \quad
{\bar X}_t = \frac{{\mathbb E}\left[
X_t\re^{\alpha X_t}\right]}{{\mathbb E}\left[ \re^{\alpha X_t}\right]}.
\label{eq:18}
\end{eqnarray}
Formula (\ref{eq:18}) shows that  $\psi''(\alpha)>0$, and thus that
the L\'evy exponent is \textit{convex} as a function of $\alpha$. 
Indeed, for any random variable  $\xi$ satisfying  $m(\alpha) := 
{\mathbb E}\,[\re^{\alpha \xi}] < \infty$ for $\alpha$ in some interval 
containing the origin, $\ln m(\alpha)$ is convex
(see, e.g., Billingsley 1995).  Now consider four 
values of $\alpha$ in $A$ such that 
$\alpha_1 < \alpha_2 \leq \alpha_3 < \alpha_4 $, 
and for some $h>0$ suppose that 
$\alpha_3=\alpha_1+h$ and $\alpha_4=\alpha_2+h$. 
Then we have:
\begin{eqnarray}
\psi(\alpha_1) +\psi(\alpha_4) >  \psi(\alpha_2) +\psi(\alpha_3). 
\label{inequality}
\end{eqnarray}
To derive this inequality we note that the convexity of $\psi(\alpha)$ implies that 
$\psi'(x+h)-\psi'(x ) >0$ for $x$ and $x+h$ in $A$. Integrating $\psi'(x+h)-\psi'(x )$ with 
respect to $x$ between $\alpha_1$ and $\alpha_2$ one obtains (\ref {inequality}). 
Then since $\psi(0) = 0$, and either $-\lambda < 0 \leq \sigma - \lambda < \sigma$ or 
$-\lambda < \sigma - \lambda \leq 0 < \sigma$, it follows from (\ref {inequality})
by letting either $h = \sigma$ or $h = \lambda$ that
\begin{eqnarray}
\psi(-\lambda) + \psi(\sigma) > \psi(\sigma-\lambda). 
\end{eqnarray}
Therefore, 
$R(\lambda, \sigma)>0$. Furthermore, the convexity of the L\'evy 
exponent implies that 
\begin{eqnarray}
\frac{\partial R(\lambda,\sigma)}{\partial\lambda}=   \psi'(\sigma-\lambda) 
-\psi'(-\lambda) \, > 0 
\quad \rm{and} \quad  
\frac{\partial R(\lambda,\sigma)}{\partial\sigma}= \psi'(\sigma) 
- \psi'(\sigma-\lambda) \, > 0.
\end{eqnarray}
Thus, $R(\lambda, \sigma)$ is  increasing with respect to both 
$\lambda$ and $\sigma$. \hfill$\Box$

\vspace{.2cm}
We observe  that the risk premium  $R(\lambda, \sigma)$ is in general 
a nonlinear function of the risk aversion (represented by $\lambda$) and the 
risk (represented by $\sigma$). This suggests that the notion of ``market price 
of risk", so common in the finance literature, is somehow linked specifically to 
models based on Brownian motion, and is not quite the right idea in the context 
of general L\'evy models. Rather, risk premium is the more useful 
notion. In Section~\ref{sec:5} we show that the only GLM leading to a bilinear risk premium
 is the GBM model. 

Properties of the risk premium can be examined further by use of the 
L\'evy-Khintchine representation for $\psi(\alpha)$, which in the case of a 
L\'evy process admitting exponential moments takes the form 
\begin{eqnarray}
\psi(\alpha)=p\alpha+\frac{1}{2}q\alpha^2+
\int_{-\infty}^\infty(\re^{\alpha x}-1-\alpha x{\mathds 1}\{|x|<1\})\, \nu(\rd x), 
\label{L-K}
\end{eqnarray}
where $p$ and 
$q>0$ are constants and $\nu(\rd x)$ is the L\'evy measure (see, e.g., 
Sato 1999, Theorem 25.17). For any measureable set $B \in {\mathds R}$, the expected 
rate at which jumps occur for which the jump size lies in the range $B$ is $\nu(B)$. It 
follows from (\ref{L-K}) that the risk premium  is given by
\begin{eqnarray}
R(\lambda,\sigma) = \half q \lambda \sigma + 
\int_{-\infty}^\infty (\re^{\sigma x}-1) (1-\re^{-\lambda x}) \, \nu(\rd x), 
\label{R L-K}
\end{eqnarray}
from which various of its properties can be deduced. In particular, the statement of 
Proposition~\ref{prop:1} can be seen to follow from the fact that the argument of the 
integrand in (\ref{R L-K}) is positive for $\sigma,\lambda>0$, and that its first derivatives 
are positive. One can also calculate the higher derivatives of the risk premium with 
respect to risk aversion and volatility, and  one deduces, for example, that in the case 
of a spectrally negative L\'evy process (downward jumps), we have $\partial_\sigma^2 
R<0$ and $\partial_\lambda^2 R>0$, and that these inequalities are reversed in the 
case of a spectrally positive process. One also observes that, providing the tails of the 
L\'evy measure are not too fat, for small values of the risk aversion and volatility the risk 
premium is approximately bilinear.

\section{L\'evy models for foreign exchange}
\label{sec:4}

\noindent When the geometric L\'evy model is extended to the case of foreign exchange, 
additional features arise that are of some interest. It is reasonable to require 
``numeraire symmetry'' 
in the sense that if, for example, the dollar price of one pound 
sterling offers a positive excess rate of return above the interest rate differential, then the 
sterling price of one dollar should offer a positive excess rate of return above the reverse 
interest rate differential. 

We examine the GBM case first, where the situation is  transparent. 
Let the dollar be the domestic currency, and the pound the foreign currency. 
Let $S_t$ denote the price of one pound in dollars, and $\tilde S_t$ the price of one 
dollar in pounds. We write  $r$ for the domestic (dollar) interest rate, and $f$ for the 
foreign (sterling) interest rate, both assumed constant. Let $\lambda$ and $\sigma$ 
be positive constants, and let the dollar pricing kernel be given by (\ref{eq:1}). Then 
the GBM model for the foreign exchange rate (the dollar price of one pound) is given by
\begin{eqnarray}
S_t = S_0\, \re^{(r-f)t}\, \re^{\lambda\sigma t} \re^{\sigma W_t- \frac{1}{2}\sigma^2 t}.
\label {brownian FX}
\end{eqnarray}
We observe that the excess rate of return above the interest rate differential $r-f$ is the 
product $\lambda \sigma$, which is positive. 
For the corresponding inverse exchange rate we obtain
\begin{eqnarray}
{\tilde S}_t = {\tilde S}_0\, \re^{(f-r)t}\, 
\re^{ (\sigma - \lambda)\sigma t}\, \re^{-\sigma W_t-\frac{1}{2}\sigma^2t} . 
\label {inverse brownian FX}
\end{eqnarray}
In the case of $\tilde S_t$, we see that the excess rate of return above the reverse interest 
rate differential $f-r$  is positive if and only if 
$\sigma > \lambda$. 
In equilibrium, we may presume that investors on both sides of the Atlantic wish to see 
the exchange rate promising a positive excess rate of return. The argument above shows 
that in a GBM model this possibility can be realised if $\sigma > \lambda$. This is the 
essence of the so-called Siegel (1972) paradox.  
 
Now let us look at the analogous situation in the context of a geometric L\'evy model.
In the case of a GLM our elementary model for the exchange rate takes the form
\begin{eqnarray}
S_t = S_0\, \re^{(r-f)t}\, \re^{R(\lambda,\sigma)t}\, \re^{\sigma X_t-t\psi(\sigma)} ,
\label{FX Levy}
\end{eqnarray}
where $\lambda$ and $\sigma$ are positive constants, and $R(\lambda,\sigma)$ is 
given by (\ref{R}), as before. Thus for the inverse exchange rate we obtain
\begin{eqnarray}
{\tilde S}_t = {\tilde S}_0 \, \re^{(f-r)t}\, \re^{{\tilde R}(\lambda,\sigma)t}\, 
\re^{-\sigma X_t-t\psi(-\sigma)},
\label {FX}
\end{eqnarray}
where ${\tilde R}(\lambda,\sigma)=-R(\lambda,\sigma)+\psi(\sigma)+\psi(-\sigma)$.
A short calculation shows that 
\begin{eqnarray}
{\tilde R}(\lambda,\sigma) = \psi(-\sigma) + \psi(\sigma-\lambda) 
- \psi(-\lambda) . 
\label{R tilde}
\end{eqnarray}
Now suppose that $\sigma > \lambda$. Then
$-\sigma < - \lambda < 0 < \sigma - \lambda$,
which by taking $h =  \sigma$ in 
equation (\ref{inequality}) implies that $\tilde R(\lambda, \sigma) > 0$. 
On the other hand, suppose that  $\sigma \leq \lambda$. 
Then since either $- \lambda \leq -\sigma \leq \sigma - \lambda \leq 0$ 
or $- \lambda \leq \sigma - \lambda  \leq - \sigma \leq 0$, it follows from (\ref{inequality}) 
by letting either $h = \sigma$ or $h =  \lambda$  that $\tilde R(\lambda, \sigma) \leq 0$. 
One thus deduces the following: 

\begin{prop}\label{prop:2}
If the volatility exceeds the risk aversion, then both {\rm (a)} the excess rate of return on 
the FX rate, and {\rm (b)} the excess rate of return on the inverse FX rate, are 
\textit{positive} in a geometric L\'evy model for foreign exchange. 
\end{prop}

One observes that the volatility of the inverse exchange rate in (\ref {FX}), as matters 
stand, appears to be negative, which is not consistent with our original characterisation 
of an investment-grade asset. We can however put the inverse exchange rate into 
``canonical" form by regarding it as being driven by the mirror process  
 $\tilde X_t = -X_t$. Then $\sigma$ is the volatility, and we can regard the 
quantity  $\tilde \lambda = \sigma - \lambda$ as being the associated risk aversion 
parameter for foreign agents. Writing $\tilde \sigma= \sigma$, and regarding 
$\tilde \sigma$ and $\tilde \lambda$ as independent variables, one can check that 
the inverse excess rate of return is monotonically increasing in both the volatility 
$\tilde \sigma$ and the foreign risk aversion factor $\tilde \lambda$. Indeed, if we 
write ${\tilde \psi}(\alpha) = \psi(-\alpha)$ for the L\'evy exponent associated with the 
mirror process, then it is an exercise to verify that 
\begin{eqnarray}
{\tilde R}(\lambda,\sigma) = \tilde \psi(\tilde \sigma) + \tilde  \psi(- \tilde \lambda) 
-\tilde \psi(\tilde \sigma- \tilde \lambda), 
\end{eqnarray}
and one sees that the form of the foreign excess rate of return, when expressed in 
terms of the relevant L\'evy exponent, is identical to that of the domestic excess rate 
of return. The requirement $\sigma > \lambda$ can be understood as an assertion 
that the foreign risk aversion is positive. Thus one might regard this constraint as a 
necessary feature of the model. 

\begin{prop}\label{prop:3}
In a geometric L\'evy  model for foreign exchange with $\sigma > \lambda > 0$, the 
excess rate of return on the inverse FX rate is  increasing with respect 
to  the independent variables $\tilde \lambda = \sigma - \lambda$ and $\tilde \sigma 
= \sigma$. 
\end{prop}

\section{Examples of Geometric L\'evy Models }
\label{sec:5}

\noindent It will be instructive to look at various explicit examples of geometric L\'evy 
models for asset prices, noting in particular the structure of the excess rate of return 
function in each case. 

\vspace{0.30cm} 
\noindent {\bf Example 1: Brownian motion}. In the case of a standard geometric 
Brownian motion model the L\'evy exponent is given by
$\psi(\alpha) = \frac{1}{2}\alpha^2$,
and hence 
\begin{eqnarray}
R(\lambda,\sigma) = \psi(\sigma) + \psi(-\lambda) - \psi(\sigma-\lambda) 
= \frac{1}{2}\sigma^2 + \frac{1}{2}\lambda^2 - 
\frac{1}{2}(\sigma-\lambda)^2 
= \sigma \lambda, 
\end{eqnarray}
which is positive. Further, for the inverse excess rate of return we have
\begin{eqnarray}
{\tilde R}(\lambda,\sigma) = \psi(-\sigma) + \psi(\sigma-\lambda) - \psi(-\lambda) 
= \frac{1}{2}\sigma^2 + \frac{1}{2}(\sigma-\lambda)^2 - 
\frac{1}{2}\lambda^2 = \sigma(\sigma-\lambda).  
\end{eqnarray}
Thus ${\tilde R}(\lambda,\sigma) $ is positive if and only if $\sigma>\lambda$. We 
observe that if one treats the quantities $\tilde \sigma = \sigma$ and 
$\tilde \lambda = \sigma-\lambda$ as independent variables, then the inverse excess 
rate of return function is increasing with respect to each. 

One can ask to what extent the bilinear form of the excess rate of return 
determines the underlying L\'evy process. Is it uniquely Brownian motion that has 
this property? If we consider expression (\ref{R}) and set $R(\lambda, \sigma) = 
\lambda \sigma$, then by taking two derivatives we deduce that $\psi''(\alpha)=1$. 
Integrating twice and imposing the condition $\psi(0) = 0$, we conclude that the
general form of the  L\'evy exponent compatible with a bilinear excess rate of return 
is 
\begin{eqnarray}
\psi(\alpha) = p \,\alpha +  \tfrac{1}{2}\alpha^2 ,
\end{eqnarray}
where $p$ is a constant, and we obtain a standard Brownian motion with drift. But the 
addition of a drift to the driving L\'evy process is irrelevant to the resulting pricing 
model in our scheme, since it cancels out in the formula for the geometric L\'evy 
martingale. Thus: 

\begin{prop}\label{prop:x}
The geometric Brownian motion model is the only geometric L\'evy model with an excess rate of return
that is bilinear in the risk aversion and the volatility. 
\end{prop}

In fact,  in any GLM the excess rate of return function is sufficient to determine the 
driving L\'evy process, up to an irrelevant drift. To establish that this is the case, we 
observe that if we differentiate each side of (\ref{R}) with respect to $\lambda$ and  
$\sigma$, and
then take the limit as $\lambda$ approaches zero,  we obtain  
\begin{eqnarray}
\psi'' (\sigma) =
\left . \frac{\partial^2 R(\lambda,\sigma)} {\partial \lambda \,\partial \sigma} \right |_{\lambda = 0}.
\end{eqnarray}
Integrating twice, and fixing the constant, we obtain the L\'evy exponent, modulo a drift.

\vspace{0.3cm} 
\noindent {\bf Example 2: Poisson process}.
Let $\{N_t\}$ be a standard Poisson process with jump rate $m>0$. Then for any nonnegative 
integer $n$ the distribution of $N_t$ 
is given by
\begin{eqnarray}
{\mathbb P}(N_t = n) =\re^{-mt} \frac{(m t)^n}{n\,!}.
\end{eqnarray}
It follows that ${\mathbb E}[X_t] = mt$, and that the L\'evy exponent is
$\psi(\alpha) = m(\re^\alpha -1)$.
The associated geometric L\'evy martingale with volatility $\alpha$ in this example 
is thus
\begin{eqnarray}
M_t = \exp\, [ \alpha N_t - mt (\re^{\alpha} -1) ] .
\end{eqnarray}
A calculation then shows that the excess rate of return function is manifestly 
positive, and increasing with respect to its arguments:
\begin{eqnarray}
R(\lambda,\sigma) = m (1-\re^{-\lambda}) ( \re^\sigma - 1).
\end{eqnarray}
We remark that since the jumps in the geometric Poisson model 
are upward, the ``risk" that an investor faces is that there may be fewer jumps than 
one hopes for. This is made evident if we combine the expressions for the 
geometric martingale and the excess rate of return function to obtain the following 
formula for the price of a non-dividend-paying asset:
\begin{eqnarray}
S_t =S_0 \exp\, [rt + \sigma N_t - mt \,\re^{-\lambda} (\re^{\sigma} -1) ] .
\end{eqnarray}
Thus, the effect of investor risk aversion is to reduce the downward drift rate in the 
compensator term by attaching the factor $\re^{-\lambda}$ to it. For the associated 
pricing kernel one has
\begin{eqnarray}
\pi_t = \exp\, [ - rt -\lambda N_t - mt (\re^{-\lambda} -1) ] ,
\end{eqnarray}
and it is an exercise to check that the product of $\pi_t$ and $S_t$ gives a geometric 
Poisson martingale with volatility $\sigma - \lambda$.
In the event that $S_t$ represents the price of a unit of foreign currency, then 
we include the foreign interest rate by setting
\begin{eqnarray}
S_t =S_0 \exp\, [(r-f)t + \sigma N_t - mt \,\re^{-\lambda} (\re^{\sigma} -1) ] .
\end{eqnarray}
For the corresponding inverse exchange rate we obtain
\begin{eqnarray}
{\tilde S}_t = {\tilde S}_0 \,\exp\, [(f-r)t +{ \tilde R(\lambda, \sigma)}- \sigma N_t + mt \, 
(1-\re^{-\sigma} ) ] ,
\end{eqnarray}
where
${ \tilde R(\lambda, \sigma)} =m(\re^{\sigma - \lambda} -1)(1 -  \re^{-\sigma} ) $,
in agreement with equation (\ref{R tilde}). If $\sigma > \lambda$ then the excess 
rate of return of the inverse exchange rate is evidently positive, and  has the 
property of being increasing with respect to  the independent 
variables $\sigma$ and $\sigma - \lambda$.

\vspace{0.3cm} 
\noindent {\bf Example 3: Compound Poisson process}. 
Let $\{N_t\}$ be a standard Poisson process with rate $m$, and let 
$\{Y_k\}_{k \in \mathbb N}$ be a collection of identical independent copies of 
a random variable $Y$ with the property that 
\begin{eqnarray}
\phi(\alpha) := {\mathbb E}\,[\re^{\alpha Y}] < \infty  
\label{phi}
\end{eqnarray}
for $\alpha$ in some connected interval $A$ containing the origin. 
Writing ${\mathds 1}\{-\}$ for the indicator function, one can check that
\begin{eqnarray}
X_t = \sum_{k=1}^{\infty} {\mathds 1}\{k \leq N_t\}\,Y_k
\end{eqnarray}
defines a L\'evy process, and that the associated L\'evy exponent is 
given by the formula
$\psi(\alpha) = m\left(\phi(\alpha) - 1\right)$.
In this example the excess rate of return function is 
\begin{eqnarray}
R(\lambda, \sigma) = m \left( \phi(\sigma) +  \phi(-\lambda) 
- \phi(\sigma-\lambda) - 1 \right),
\end{eqnarray}
and the fact that it is positive and is bi-monotonic in its arguments is evident as 
a consequence of the convexity of $\phi(\alpha)$, which  follows
from (\ref{phi}). If $S_t$ represents the price of a unit of foreign currency, then 
the resulting geometric L\'evy model for the exchange rate is
\begin{eqnarray}
S_t =S_0 \exp\, [(r-f)t + \sigma X_t + mt \,( \phi(-\lambda) - \phi(\sigma-\lambda)) ] .
\end{eqnarray}

\vspace{0.3cm} 
\noindent {\bf Example 4: Jump diffusion  process}. 
This example is a generalisation of the Merton (1976) jump 
diffusion  model.  The driver is a vector L\'evy process, one component being 
a standard Brownian motion, and the other a compound Poisson process with 
normally distributed jumps. Write $\lambda$ and $\sigma$ for the risk aversion 
and volatility of the Brownian component, and $\beta$ and $\theta$  for 
the risk aversion and volatility of the jump component. Let $Y$ be 
normally-distributed with mean zero and variance one. Then 
$\phi(\alpha) = \exp ( \frac{1}{2} \alpha^2 ) $, and the excess rate of return function 
is of the form
\begin{eqnarray}
R(\lambda, \sigma, \beta, \theta) = \lambda \sigma + m \left( \re^{  \frac{1}{2} 
\theta^2} + \re^{  \frac{1}{2} \beta^2} - \re^{  \frac{1}{2}( \theta-\beta)^2} -1 \right),
\end{eqnarray}
which is positive and is monotonic in each of the variables. 
In Merton (1976) a key notion used to price options is the idea that  there is no 
risk premium offered by the market in connection with idiosyncratic firm-specific risk. 
Merton assumes that jump risk is purely idiosyncratic and can be 
diversified away by holding a suitably broad portfolio. Merton argues that since 
the risk can be diversified away, the market awards no risk premium to investors 
who hold such assets. From a modern point of view, the assumption that jump 
risk is necessarily idiosyncratic is questionable: this is one of the lessons of the 2008  
credit crisis. In our version of Merton's model, however, jump risk is being priced.


\vspace{0.3cm} 
\noindent {\bf Example  5: Gamma process}. 
By a standard gamma process with growth rate $m$, we mean a 
process $\{\gamma_t\}$ that has gamma-distributed 
stationary and independent increments, and satisfies 
${\mathbb E}\,[\gamma_t] = mt $ and ${\rm Var}\,[\gamma_t] = mt. $
The density of $\gamma_t$ is  given by
\begin{eqnarray}
{\mathbb P}(\gamma_t \in \rd x)=
{\mathds 1}\{x>0\} \frac{x^{mt-1}\re^{-x}} {\Gamma[mt]} \, \rd x, 
\label{eq:gamma}
\end{eqnarray}
where $\Gamma[a]$ denotes the gamma
function.
The identity
$\Gamma[a+1]=a\Gamma[a]$ implies that the mean of $\gamma_t$ is $mt$, thus
justifying the interpretation of $m$ as the growth rate.
The associated moment generating function is 
\begin{eqnarray}
{\mathbb E}[\re^{\alpha \gamma_t}] =(1-\alpha)^{-mt}
= \re^{-mt\ln(1-\alpha)},
\end{eqnarray}
and hence the L\'evy exponent is 
\begin{eqnarray}
\psi(\alpha) = -m \ln(1-\alpha),
\label{psi gamma}
\end{eqnarray}
which is well-defined for $\alpha < 1$. 
For some applications it is useful to consider the two-parameter family of
so-called scaled gamma processes. By a scaled gamma process with 
growth rate $\mu$ and variance rate $\nu^2$ we mean a process 
$\{{\mathit\Gamma}_t\}_{0 \leq t<\infty}$ with stationary and independent 
increments such that ${\mathit\Gamma}_0=0$
and such that ${\mathit\Gamma}_t$ has a gamma distribution with mean $\mu t$
and variance $\nu^2t$, where $\mu$ and $\nu$ are parameters.
Setting $m=\mu^2/\nu^2$ and $\kappa=\nu^2/\mu$, one finds that $\mu=
\kappa m$ and $\nu^2=\kappa^2 m$. One can thus think of $m$ as a
standardised growth rate, and $\kappa$ as a scale parameter. The
density of ${\mathit\Gamma}_t$ is given by
\begin{eqnarray}
{\mathbb P}({\mathit\Gamma}_t \in \rd x) = {\mathds 1}\{x>0\}
\frac{\kappa ^{-mt} x^{mt-1} \re^{-x/\kappa }}{\Gamma[mt]} \, \rd x .
\label{eq:ww2}
\end{eqnarray}
For fixed $t$ the product $mt$ is the so-called shape parameter of the 
distribution of the random variable ${\mathit\Gamma}_t$. If $\{\gamma_t\}$ is a 
standard gamma process with growth rate $m$, then the process 
$\{{\mathit\Gamma}_t\}$ defined by ${\mathit\Gamma}_t  = \kappa  \gamma_t $ is 
evidently a scaled gamma process with standardised growth rate $m$ 
and scale parameter $\kappa$, and for its moment generating function 
we have
\begin{eqnarray}
{\mathbb E}[\re^{\alpha {\mathit\Gamma}_t}] =(1-\alpha \kappa)^{-mt}.
\label{scaled gamma psi}
\end{eqnarray}
See Dufresne \textit{et al}.~(1991),  Dickson \& Waters (1993), 
Madan \textit{et al}.~(1998), Baxter (2007), Yor (2007), and Brody 
\textit{et al}.~(2008) for various aspects of the theory of the gamma process.

Now let  $\{\gamma_t\}$ be a standard gamma process with growth 
rate $m$, and let $\sigma$ be a constant such that $0<\sigma<1$. 
The associated geometric L\'evy martingale takes the form
\begin{eqnarray}
M_t = (1-\sigma)^{mt} \re^{\sigma \gamma_t}.
\end{eqnarray}
The jumps are upward, and the compensator is a deterministic decreasing 
process. If follows from (\ref{R}) and (\ref{psi gamma}) that the excess rate 
of return function is of the form
\begin{eqnarray}
R(\lambda,\sigma) = m \ln \frac{1-\sigma+\lambda}{(1-\sigma)(1+ \lambda)} . 
\label{R gamma}
\end{eqnarray}
If one takes the difference between the numerator and the denominator in 
the argument of the logarithm in (\ref{R gamma}), the result is
$(1-\sigma+\lambda) - (1-\sigma)(1+\lambda) = \sigma \lambda$ ,
which is positive. It follows that $R(\lambda,\sigma)>0$. 
By (\ref{FX Levy}), the
corresponding model for the foreign exchange rate is
\begin{eqnarray}
S_t =S_0\, \re^{(r-f)t}\left(1-\frac{\sigma}{1+\lambda}\right)^{mt}
\re^{\,\sigma \, \gamma_t}.
\label{gamma FX rate}
\end{eqnarray}
Variants of the geometric gamma model appear in Heston (1993), 
Gerber \& Shiu (1994), and Chan (1999). We observe that the effect of 
risk aversion is to reduce the rate at which the compensator decreases, 
thus encouraging investors who might otherwise be concerned over the 
possibility of an insufficient rise in the underlying gamma process. 
For small $\lambda$ and  $\sigma$ the risk premium is given approximately 
by $ \lambda \, \sigma $.

As for the excess rate of return associated with the inverse exchange rate, 
by use of (\ref{R tilde}) we obtain the following expression:
\begin{eqnarray}
{\tilde R}(\lambda,\sigma) = m \ln \frac{1+\lambda}{(1+\sigma)
(1-\sigma+\lambda)} .
\end{eqnarray}
We observe that  
$(1+\lambda) - (1+\sigma)(1-\sigma+\lambda) = \sigma (\sigma-\lambda)$,
from which it follows that ${\tilde R}(\lambda,\sigma) >0$ if and only if 
$\sigma>\lambda$. Numeraire symmetry thus imposes a bound on the risk 
aversion factor and we have $0< \lambda<\sigma<1$. Inverting 
(\ref{gamma FX rate}), and writing $\tilde \sigma = \sigma$ and 
$\tilde \lambda = \sigma - \lambda$, we find that 
 \begin{eqnarray}
\tilde S_t =\tilde S_0\, \re^{(f-r)t}\left(1+\frac{\tilde \sigma}
{1-\tilde \lambda}\right)^{mt}\re^{-\, \tilde \sigma  \,\gamma_t}.
\end{eqnarray}
Thus $\tilde S_t$ is driven by a negative gamma process, which jumps 
downward, and the compensator is a deterministic increasing process. 
Since $0<\tilde \lambda <1$, the effect of foreign risk aversion is to 
increase the rate at which the compensator increases. 
For small values of $\lambda$ and $\sigma$  the inverse excess 
rate of return is given approximately by ${\tilde\lambda}\, {\tilde\sigma}$.

\vspace{0.3cm} 
\noindent {\bf Example 6: Variance gamma process}. 
It will be convenient first to discuss the symmetric variance 
gamma (VG) process. 
This is the process considered by Madan \& Senata (1990) and 
Madan \& Milne (1991). Then in the next example we discuss the more 
general asymmetric or ``drifted" VG process of Madan \textit{et al}.~(1998). 
Both of these processes are of interest from a mathematical perspective 
and as a basis for financial modelling. There is a further extension 
of the model, due to Carr \textit{et al}.~(2002), which will not be discussed here. 
The VG model relies on the use of a 
gamma process as a subordinator. Thus we begin with a standard gamma 
process $\{\gamma_t\}$ with rate $m$, 
and give it the dimensionality of time by dividing it by $m$. 
In this way we define a scaled 
gamma process $\{{\mathit\Gamma}_t\}$ by setting
${\mathit\Gamma}_t = m^{-1} \gamma_t$,
and we observe that ${\mathbb E}\,[{\mathit\Gamma}_t] = t $. 
We call $\{{\mathit\Gamma}_t\}$ 
a standard gamma subordinator. The symmetric VG process 
$\{V_t\}$, with parameter $m$, is defined by letting $\{W_t\}$ be a standard 
Brownian motion and setting $V_t = W_{{\mathit\Gamma}_t }$. The 
associated moment 
generating function is thus
\begin{eqnarray}
{\mathbb E}\left[\exp ({\alpha V_t})\right ]= 
{\mathbb E}\left[\exp ({\alpha W_{{\mathit\Gamma}_t}})\right ]=
{\mathbb E}\,\left[\exp \Big({\frac{1}{2} \alpha^2 {\mathit\Gamma}_t\Big)}\right] =
\left(1-\frac {\alpha^2}{2m}\right)^{-mt},
\end{eqnarray}
which is defined for $\alpha^2 < 2m$. Clearly  $\alpha$ must have units of 
inverse square-root time, since $m$ has units of inverse time; but this is 
consistent with the fact that $V_t$  has units of square-root time, like 
the Wiener process. The associated L\'evy exponent is
\begin{eqnarray}
\psi(\alpha) = -m \ln \left(1-\frac {\alpha^2}{2m}\right),
\end{eqnarray}
and one can check the convexity of $\psi(\alpha)$ in this example by 
observing that 
\begin{eqnarray}
\psi''(\alpha) = m \left(1-\frac {\alpha^2}{2m}\right)^{-2}.
\label{VG psi''}
\end{eqnarray}
As a consequence the geometric L\'evy martingale in the symmetric VG case 
takes the form 
\begin{eqnarray}
M_t =  \left(1-\frac {\alpha^2}{2m}\right)^{mt}\exp\, \left({\alpha 
W_{{\mathit\Gamma}_t}}\right), 
\end{eqnarray}
and the excess rate of return function, which  is positive and monotonic, 
is 
\begin{eqnarray}
R(\lambda,\sigma) = 
m \ln \left[  \left( 1 -\frac {(\sigma - \lambda)^2}{2m} \right)
 \left(1 -\frac { \lambda^2}{2m} \right)^{-1} 
  \left(1 -\frac { \sigma^2}{2m} \right)^{-1} \right].
\end{eqnarray}
The corresponding VG foreign exchange rate is thus given by
\begin{eqnarray}
S_t = S_0\, \re^{(r-f)t}  \left( 1 -\frac {(\sigma - \lambda)^2}{2m}\right)^{mt} 
\left({1 -\frac { \lambda^2}{2m}}\right)^{-mt} \exp\, \left({\sigma 
W_{{\mathit\Gamma}_t}}\right).
\end{eqnarray}
We remark that in the case of the VG model one finds by use of (\ref{VG psi''}) 
that the risk premium satisfies $\partial_\sigma^2 R>0$ iff  $\sigma > | \sigma - 
\lambda |$,  and $\partial_\lambda^2 R>0$ iff $\lambda > |\sigma - \lambda |$.

A well known alternative characterisation of the VG process is 
as follows. Let $\{\gamma^1_t\}$ and $\{\gamma^2_t\}$ be a pair of 
independent standard gamma processes, each with rate $m$. Then the process 
defined by the difference between these two processes has both upward and 
downward jumps, and is symmetrical about the origin in distribution, with 
mean zero. If we normalise the difference by setting
\begin{eqnarray}
V_t =\frac {1}{\sqrt{2m}} \left( \gamma^1_t - \gamma^2_t \,\right),
\label{VG}
\end{eqnarray}
then it is easy to check that the variance of $V_t$ is $t$, and so we get a pure 
jump process that has some properties in common with Brownian motion. Indeed 
if we consider the moment generating function we find by virtue of the 
independence of the two gamma processes that
\begin{eqnarray}
{\mathbb E}\left[\exp ({\alpha V_t})\right ]&=&
{\mathbb E} \left[ \exp \left(\alpha \frac {1}{\sqrt{2m}} \left( \gamma^1_t 
- \gamma^2_t \,\right) \right) \right ] 
\nonumber \\ &=& \left(1-\frac {\alpha}{\sqrt {2m}}\right)^{-mt} \left(1
+\frac {\alpha}{\sqrt {2m}}\right)^{-mt} 
\nonumber \\
&=& \left(1-\frac {\alpha^2}{2m}\right)^{-mt},
 \end{eqnarray}
and it is evident (Madan \& Senata 1990) that  (\ref{VG}) has the law of a VG process. 
For large values of $m$, 
the distribution of $V_t$ is approximately Gaussian. In particular, we have:
\begin{eqnarray}
\lim_{m \to \infty} {\mathbb E}\left[\exp ({\alpha V_t})\right ]= \exp 
\left(\frac {1}{2} \alpha^2 t \right).
\end{eqnarray}

\vspace{0.3cm} 
\noindent {\bf Example 7: Asymmetric VG process}. 
The representation of the VG process as the normalised difference between 
two independent gamma processes suggests two generalisations. One is 
that of Madan \textit{et al}.~(1998), where we consider an asymmetric 
difference between two independent standard gamma processes. Thus writing 
\begin{eqnarray}
U_t =\kappa_1 \gamma^1_t - \kappa_2 \gamma^2_t ,
\end{eqnarray}
where $\kappa_1$ and $\kappa_2$ are nonnegative constants, a calculation of 
the respective moment generating functions shows that $U_t$ is identical in law 
to a ``drifted" VG process of the form 
\begin{eqnarray}
U_t =\mu {\mathit\Gamma}_t + \sigma W_{{\mathit\Gamma}_t},
\end{eqnarray}
where $\mu$ and $\sigma$ are constants. The relationship between $\mu$, 
$\sigma$, $\kappa_1$, $\kappa_2$, and $m$ is given by 
$\mu = m (\kappa_1 - \kappa_2)$ and 
$ \sigma^2 = 2m \kappa_1  \kappa_2 \,$, together with
\begin{eqnarray}
\kappa_1=   \frac {1}{2m} \left( \mu + \sqrt{\mu^2 + 2m \sigma^2} \right)    
\quad {\rm and} \quad
\kappa_2=   \frac {1}{2m} \left( -\mu + \sqrt{\mu^2 + 2m \sigma^2} \right).   
\end{eqnarray}
The L\'evy exponent $\psi(\alpha) = -m \ln \left(1-( \kappa_1 -  \kappa_2)\, 
\alpha - \kappa_1 \kappa_2\, \alpha^2  \right)$, which can be worked out by use of 
(\ref{scaled gamma psi}), can be equivalently written in the form
\begin{eqnarray}
\psi(\alpha) = -m \ln \left(1-\frac{\mu}{m}\, \alpha - \frac{\sigma^2}{2m}\, 
\alpha^2  \right),
\end{eqnarray}
where the range of $\alpha$ is $-1/\kappa_2 < \alpha < 1/\kappa_1$. 
It is straightforward to write down the associated excess rate 
of return function, and the corresponding expression for an asset price. In this 
example there is a single risk aversion factor. 

On the other hand, one can also envisage the situation where the two gamma 
drivers are regarded as separate sources of risk, each being assessed 
independently by the market. This situation arises in instances where investors 
are for some reason more worried about downward jumps than upward ones. 
More specifically, let us suppose that investors are more concerned about 
excessive losses than about insufficient gains. It is said that studies in 
behavioural finance suggest that this may actually be the case.  One can 
model such a situation rigorously by introducing an asymmetric pricing kernel 
of the form 
\begin{eqnarray}
\pi_t = \re^{-rt}\,(1-\lambda_1)^{mt}  \,(1+\lambda_2)^{mt} \re^{-\lambda_1 
\gamma^1_t} \re^{\lambda_2 \gamma^2_t} , 
 \end{eqnarray}
 and an asset price process  of the form 
 \begin{eqnarray}
S_t =S_0\, \re^{rt}\left(1-\frac{\sigma_1}{1+\lambda_1}\right)^{mt}
\left(1+\frac{\sigma_2}{1-\lambda_2}\right)^{mt}\re^{\sigma_1 \gamma^1_t}
\re^{-\sigma_2 \gamma^2_t}.
\end{eqnarray}
Thus we have separate risk aversion factors for the upward jumps and the 
downward jumps. It is interesting to observe that in the case of  ``behavioural 
asymmetry" both the asset price and the pricing kernel are driven by extended 
VG processes---but there are two distinct such processes, one driving the 
pricing kernel, and the other driving the asset price. Indeed, the pricing kernel 
is driven by $\lambda_1 \gamma^1_t - \lambda_2 \gamma^2_t $, whereas the 
asset price is driven by $\sigma_1 \gamma^1_t - \sigma_2 \gamma^2_t $. 
These processes are synchronised in the sense that the times of their upward 
and downward jumps coincide and the magnitudes are proportional for a given 
jump type. Alternatively, we can model the two driving processes as different 
linear combinations of ${\mathit\Gamma}_t$ and $W_{{\mathit\Gamma}_t}$.

\vspace{0.3cm} 
\noindent {\bf Example 8: Negative binomial process}. This process is a species 
of compound Poisson process, and can be viewed as a special case of 
Example 3. It is nevertheless of considerable interest in its own right. See, for 
example, Kozubowski \& Krzysztof (2009) for a general overview. The negative 
binomial process has a L\'evy exponent of the form 
\begin{eqnarray}
\psi(\alpha) = m \ln\left( \frac{1-q}{\,1-q\re^{\alpha}}\right),
\label {negative binomial}
\end{eqnarray}
where $0<q<1$, and $m>0$ is a rate parameter. A short calculation shows that
\begin{eqnarray}
R(\lambda,\sigma) = m\, \ln  \left[ \frac{(1-q)(1-q\re^{\sigma-\lambda})}{(1-q\re^\sigma)
(1-q\re^{-\lambda})} \right] .
\end{eqnarray}
To see explicitly that this is positive we take the difference between the numerator 
and the denominator inside the logarithm to obtain $q(\re^\sigma-1) (1-\re^{-\lambda})$,
which clearly is positive since $\sigma,\lambda>0$. It follows that the argument of the 
logarithm is larger than one, and we have $R(\lambda,\sigma)>0$. For the inverse 
excess rate of return  we obtain 
\begin{eqnarray}
{\tilde R}(\lambda,\sigma) = m \, \ln  \left[ \frac{(1-q)(1-q\re^{-\lambda})}
{(1-q\re^{-\sigma})
(1-q\re^{\sigma-\lambda})} \right] . \label{eq:zz1}
\end{eqnarray}
To analyse the positivity of ${\tilde R}(\lambda,\sigma)$ we again take the 
difference between the numerator and the denominator in the argument of the 
logarithm in (\ref{eq:zz1}). We get $q ( \re^{\sigma - \lambda} - 1) (1-\re^{-\sigma})$,
which is positive only if $\sigma>\lambda$.

The jumps of the negative binomial process are positive integers.  There are two 
distinct representations for the process. The first of these takes the form of a compound 
Poisson process with the following characteristics: (a) the jump sizes have a so-called 
logarithmic distribution, given in the notation of Example 3 by
\begin{eqnarray}
{\mathbb P}(Y= n) = - \frac {1}{\ln (1-q)} \,\,\frac { 1 } { n } \, q^n\, ; 
\end{eqnarray}
and (b) the intensity $\mu$ of the underlying Poisson process is of the form $ \mu = 
-m \ln (1 -q)$. A straightforward calculation of the moment generating function of $Y$ 
gives
\begin{eqnarray}
\phi(\alpha) = \frac { \ln (1-q \re^\alpha) } { \ln (1-q)} .
\end{eqnarray}
By the general theory of the compound Poisson process we know that 
$\psi (\alpha) = \mu (\phi (\alpha) -1)$, 
which immediately leads to the L\'evy exponent (\ref {negative binomial}). If we write 
$X_t$ for the value of the negative binomial process at time $t$, we find that its 
probability mass function is given by 
\begin{eqnarray}
{\mathbb P}(X_t= k) =  \frac{\Gamma(k+mt)}{\Gamma(mt)\Gamma(k+1)} q^k (1-q)^{mt}, 
\end{eqnarray}
which is the negative binomial distribution; and for the geometric L\'evy martingale we have 
\begin{eqnarray}
M_t = \left( \frac{\,\,1-q\re^{\alpha}}{1-q}\right)^{mt} \re^{\alpha X_t}.
\end{eqnarray}

The second representation of the negative binomial process is reminiscent of the
theory of the VG process. We take a standard Poisson process, with intensity 
\begin{eqnarray}
\Lambda = m\,\frac {q}{(1-q)}, 
\label{lamba}
\end{eqnarray}
where $0<q<1$ and $m>0$ as before, and subordinate it with a gamma subordinator 
$\{{\mathit\Gamma}_t \}$ with standardised rate parameter $m$. Thus the expectation of  
${\mathit\Gamma}_t $ is $t$, and its variance is $t/m$, as in the theory of the VG process.  The 
associated moment generating function is thus
\begin{eqnarray}
{\mathbb E}\left[\exp ({\alpha N_{{\mathit\Gamma}_t}})\right ] =
{\mathbb E}\,[\exp ({\Lambda (\re^{\alpha} - 1) {\mathit\Gamma}_t)}] = \left(1-\frac {\Lambda 
(\re^{\alpha} -1 )}{m}\right)^{-mt}
= \left( \frac{1-q}{\,1-q\re^{\alpha}}\right)^{mt},
\end{eqnarray}
by virtue of the chosen intensity (\ref{lamba}), and we are lead directly to the L\'evy 
exponent (\ref {negative binomial}). 
\vspace{0.35cm}

It is natural to ask in the context of these various examples what information can be extracted 
(or ``implied") about the values of model parameters when one is given option prices. 
In the case of the GBM model, for example, it is known that one can infer the value of 
volatility $\sigma$, but that the option price is independent of risk aversion $\lambda$. This can be checked directly by working out the price of a call option with strike $K$ and expiry $T$
by inserting (\ref{eq:1}) and  (\ref{eq:2}) into the valuation formula 
$C_0 = {\mathbb E} \, [ \pi_T\, (S_T - K)^+] $.
In a general geometric L\'evy model, this is no longer the case: option prices depend on 
both the risk aversion and the volatility. Indeed, a variety of different situations can arise, 
each with its own character.  Thus in the 
Poisson model, there are  two nontrivial model parameters---the risk aversion, and the 
jump rate $m$ (the volatility is easily determined by observation of the price process); 
and a calculation shows that option prices depend on $m \re^{- \lambda}$, but not on 
$m$ or $\lambda$ separately. Thus if we can estimate the value of the actual jump rate 
$m$ by observations of the asset price, then $\lambda$ can be inferred from option 
prices. In the case of the gamma model, there are three nontrivial model 
parameters---the risk aversion, the volatility, and the jump rate. A calculation shows that 
option prices depend on $m$ and on $\sigma / (1 + \lambda)$, but not on  $\sigma$ and  
$\lambda$ separately, so neither $\lambda$ nor $\sigma$ can be determined exactly from 
option prices.

\section{Dividend paying assets}
\label{sec:6}

\noindent Thus far we have considered the case of non-dividend paying 
assets. From a conceptual point of view it is better, however,  to think of an asset 
price as being determined by the dividend stream or cash flow produced by the 
asset. Hence with the inclusion of dividends the pricing model is characterised 
by:  (a) a pricing kernel $\{\pi_t\}$; and (b) the dividend stream $\{D_t\}$ generated by 
the asset. The value of the asset at time $t$ is regarded as a \textit{derived} quantity 
which can be worked out by means of the fundamental relation 
\begin{eqnarray}
S_t = \frac{1}{\pi_t} \, {\mathbb E}_t\left[ \int_t^\infty \pi_s D_s \rd s \right]. 
\label{valuation}
\end{eqnarray}
With this in mind, let us consider how one extends the GBM 
model when dividends are included. The answer to this problem is well known, but 
rather than assuming the conclusion we shall derive it from first principles by modelling 
the dividend stream and the pricing kernel, and working out the resulting price process 
for the asset. For a typical investment asset in the GBM situation we model the dividend 
stream by  setting
\begin{eqnarray}
D_t = D_0 \re^{\gamma t} \re^{\sigma W_t - \frac{1}{2}\sigma^2t}, 
\label{dividend}
\end{eqnarray}
where $D_0$ is the initial rate at which dividends are paid, $\gamma$ is the 
growth rate of the dividend, and the constant $\sigma>0$ characterises the volatility of 
the dividend rate. We shall assume that the pricing kernel is of the form (\ref{eq:1}). 
Substituting (\ref{eq:1}) and (\ref{dividend}) into equation  (\ref{valuation}) and performing a short calculation 
under the assumption that $r+\lambda\sigma>\gamma$, we deduce that 
\begin{eqnarray}
S_t = \frac{1}{r+\lambda\sigma-\gamma} \, D_0 \,
\re^{\gamma t+\sigma W_t - \frac{1}{2}\sigma^2t}. 
\label{eq:4}
\end{eqnarray}
Thus we obtain a stochastic generalisation of the Gordon (1959) growth model, and 
at time zero we have the following valuation formula:
\begin{eqnarray}
S_0= \frac{D_0}{r+\lambda\sigma-\gamma}.
\end{eqnarray}
We observe that an increase in risk aversion has the effect of lowering the asset price, 
everything else being the same. Defining the proportional dividend rate by setting
$\delta=r+\lambda\sigma-\gamma$,
which by 
assumption is positive, we are able to deduce the relation 
$D_t=\delta S_t$, and we find that the asset 
price process is given by 
\begin{eqnarray}
S_t = S_0\, \re^{(r-\delta+\lambda\sigma)t}\, \re^{\sigma W_t - \frac{1}{2}\sigma^2t}. 
\label{eq:5}
\end{eqnarray}
As we stated earlier, the resulting expressions for $\{S_t\}$ and $\{D_t\}$ are of course 
familiar:  the point is that we derive these formulae here rather than assume them.

In the situation of a geometric L\'evy model it is remarkable that essentially the same 
line of argument carries through.
Thus we assume a pricing kernel of the form (\ref{levy pricing kernel}),
and a volatile dividend stream of the form
\begin{eqnarray}
D_t = D_0 \, \re^{\gamma t}\, \re^{\sigma X_t-t\psi(\sigma)}.
\end{eqnarray}
It is then an exercise to check that the fundamental relation gives
\begin{eqnarray}
S_t = S_0 \, \re^{(r-\delta)t}\, \re^{R(\lambda,\sigma)t}\, \re^{\sigma X_t-t\psi(\sigma)},
\end{eqnarray}
and a proportional dividend flow $D_t=\delta S_t$, where
$\delta=r+R(\lambda, \sigma)-\gamma$.
The resulting initial valuation formula is 
\begin{eqnarray}
S_0= \frac{D_0}{r+R(\lambda,\sigma)-\gamma}\,.
\end{eqnarray}
This  relation ties together in the context of a general geometric 
L\'evy model the values of the initial asset price, the initial dividend rate, the interest 
rate, the risk aversion level, the dividend volatility, and the dividend growth rate.

\section{Multi-factor models with predictable volatility}
\label{sec:7}

\noindent There are several reasons for extending the analysis to higher dimensions. 
First, we would like  to consider models for a market consisting of a number 
of different assets. Second, even in the consideration of a single asset, it is 
natural to introduce the additional complexity of a higher-dimensional process 
to describe its dynamics. Both situations are familiar in the context of Brownian 
motion driven models. It seems to be advantageous to envisage 
the entire market as being driven by a \textit{single} higher-dimensional L\'evy 
process. We can use essentially the same 
notation as in the one-dimensional case. Now $\{{X}_t\}$ is understood to be 
a vector L\'evy process. For the L\'evy exponent we still have (\ref{cumulant}), but now $\alpha$ is understood to be a vector, and 
there is an implicit inner product between $\alpha$ and $X_t$ in the exponent 
on the left hand side of (\ref{cumulant}). The L\'evy exponent is 
a function of the $n$ components of $\alpha$. 

The model is thus determined as in Section~\ref{sec:I}, with the assumption that 
the market filtration is generated by a vector L\'evy process, and with the 
specification of (i) a pricing kernel, and (ii) a collection of investment grade 
assets, driven collectively by $\{{X}_t\}$. We are accustomed, in the 
multi-dimensional Brownian case, to regard such a higher dimensional driver 
as being built from a set of  independent drivers that can be isolated after a 
suitable linear transformation. No such simplification is readily at hand for a 
general vector L\'evy process. Nevertheless for applications it is useful to 
consider the case where the components are assumed to be independent. 
This encompasses a large class of models, including  the higher-dimensional 
Brownian motion models. 

The pricing kernel is a process of the form (\ref{levy pricing 
kernel}) where $\lambda$ is now understood to be a vector risk aversion factor. 
If the components of the vector L\'evy process are independent, the L\'evy 
exponent separates into a sum of terms, one for each component of its 
argument, each term being the marginal L\'evy exponent associated with one 
of the risk factors. Next we introduce a set of investment-grade assets, each 
of the form (\ref{asset price}) for some choice of the vector volatility $\sigma$. 
We  require that $\lambda$ and $\sigma$ are  ``positive" vectors---that is, they 
belong to the cone of vectors with the property that all components are nonnegative 
and at least one component is positive. For a 
generic asset the excess rate of return  takes the form (\ref{R}), only 
now the arguments are understood to be vectors. When the L\'evy process 
has independent components, the excess rate of return  separates 
into a sum of terms, each being the excess rate of return associated with one 
of the components. In that case, we see by use of the arguments presented 
earlier that each term is nonnegative, and at least one is positive; as a 
consequence the total 
excess rate of return function is positive. Similarly one sees 
that in the case of independent components the excess rate of return is 
increasing with respect to the individual components of the 
risk aversion vector and the volatility vector. Thus we obtain the following result. 

\begin{prop}\label{prop:4}
The excess rate of return function 
$R(\lambda, \sigma)$ in a multi-dimensional geometric L\'evy model with 
independent L\'evy drivers is positive, and is increasing with respect to each 
of the components of the risk aversion vector $\lambda$ and the volatility vector 
$\sigma$. 
\end{prop}

\indent In the discussion so far, we have assumed for simplicity that the interest 
rate, the risk aversion, and the asset price volatilities are constant. Indeed, as 
with many financial models, various characteristic features of the model are already 
present under the assumption of constant coefficients; but for practical 
applications, and to take the theory further, we need to relax this condition. 
Thus in the case of geometric Brownian motion models we consider the situation 
where the risk aversion 
$\{\lambda_t\}$ and volatility $\{\sigma_t\}$ are adapted vector-valued 
processes, and are chosen in such a way as to ensure that the process 
defined by   the expression
\begin{eqnarray}
M_t = \exp \left({\int_0^t \alpha_s\, {\rm d}W_s-\tfrac{1}{2}\int_0^t\alpha_s^{\,2}
\,{\rm d}s}\right)
\end{eqnarray}
is an  $\{{\mathcal F}_t\}$-martingale for $\alpha_t = -\lambda_t$, $\alpha_t = 
\sigma_t$, and $\alpha_t = \sigma_t -\lambda_t\,$. It suffices  that $\{\alpha_t\}$ 
should be bounded. More generally, we consider the situation where $\{X_t\}$ 
is a L\'evy process with exponential moments, 
$\{{\mathcal F}_t\}$ is the associated filtration, and $\{\alpha_t\}$ is 
a \textit{predictable} process, adapted to $\{{\mathcal F}_t\}$, chosen in such a way 
that $\alpha_t\in A$ for $t\geq0$ and that the local martingale defined by
\begin{eqnarray}
M_t = \exp \left({\int_0^t \alpha_s {\rm d}X_s-\int_0^t\psi(\alpha_s){\rm d}s}\right)
\end{eqnarray}
is a martingale. If a predictable vector process $\{\alpha_t\}$ satisfies these 
conditions then we say it is admissible. Thus we consider a market model of the 
following form. Let the exogenously specified short rate process $\{r_t\}$ be adapted to 
$\{{\mathcal F}_t\}$, and be such that the unit-initialised money market account 
\begin{eqnarray}
B_t = \exp \left({\int_0^t r_s \,{\rm d}s} \right)
\label {money market account}
\end{eqnarray}
is finite almost surely for $t>0$. Let the $\{{\mathcal F}_t\}$-adapted vector risk 
aversion and volatility processes $\{\lambda_t\}$ and $\{\sigma_t\}$ be positive, 
and be such that the processes $\{-\lambda_t\}$, $\{\sigma_t\}$, and 
$\{\sigma_t -\lambda_t\}$ are admissible in the sense described above. The 
pricing kernel is taken to be of the form 
\begin{eqnarray}
\pi_t = \exp \left({-\int_0^t r_s\, {\rm d}s -\int_0^t \lambda_s \, {\rm d}X_s
-\int_0^t\psi(-\lambda_s)\,{\rm d}s}\right).
 \label{general pricing kernel}
\end{eqnarray}
The corresponding expression for the price of a typical non-dividend-paying 
asset is then
\begin{eqnarray}
S_t = S_0 \exp \left({\int_0^t r_s \,{\rm d}s + \int_0^t R(\lambda_s,\sigma_s)\, 
{\rm d}s + \int_0^t \sigma_s \,{\rm d}X_s-\int_0^t\psi(\sigma_s)\,{\rm d}s}\right),
\label{general asset}
\end{eqnarray}
where $R(\lambda,\sigma)$ is the excess rate of return function associated 
with the given L\'evy exponent. Clearly both $\{\pi_t B_t\}$ and $\{\pi_t S_t\}$ 
are martingales. It should also be evident that the following statement holds, 
which is to be understood as an expression of the fact that  the asset offers 
a rate of return greater than the interest rate.

\begin{prop}\label{prop:5}
Let the asset price in a model with predictable 
volatility and risk aversion, driven by a vector of independent L\'evy processes,  
be given by {\rm (\ref{general asset})}, and define the money market account by 
{\rm(\ref {money market account})}. Then the asset price, expressed in units of 
the money market account, is a submartingale.
\end{prop}

\noindent In the literature, stochastic volatility models are often introduced by the 
method of a time change. In the context of simple parametric volatility this 
amounts to the observation that if $\{W_t\}$ is a standard Brownian motion, 
then the processes defined by $Z_t=\sigma W_t$ and $Z'_t=W_{\sigma^2 t}$ have 
the same law. Thus instead of introducing the volatility as a coefficient measuring 
the sensitivity of the asset price to the underlying Brownian motion, it is 
introduced by ``speeding up''( or ``slowing down'') the Brownian motion: 
the effect is equivalent. In the case of jump processes, the two transformations 
are clearly inequivalent; thus, for example, in the case of the Poisson process 
the effect of scaling the process (magnifying the the jumps) is quite different from 
that of scaling the time (speeding up the arrival of jumps). It seems that in the 
general situation one wishes to consider both effects. Thus if the market is 
driven by a vector L\'evy process $\{X_t\}$, then we introduce a vector 
of sensitivity parameters $\sigma$, as well as a time dilation factor $c$, and 
let a typical asset be driven by the process $\{\sigma X_{ct}\}$. Again, we think 
of the entire market as being driven by a single vector L\'evy process so it is 
consistent that there is but a single overall time change for that process. On 
the other hand if the time change is random then we introduce a subordinator 
$\{c_t\}$ and the time-changed process is given by $\{\sigma X_{c_t}\}$. If the 
subordinator is itself a L\'evy process then one stays within the category of 
models already under consideration---thus the VG process can be obtained 
by subordinating Brownian motion with a gamma process, and the negative 
binomial process can be obtained by subordinating a Poisson process with a 
gamma process, as we have seen. 

\section{Concluding remarks}
\label{sec:8}

\noindent Geometric L\'evy models have a surprisingly wide range of
desirable properties. As we have seen, once suitable inequalities are imposed 
on the volatility and risk aversion parameters, the convexity of the L\'evy 
exponent ensures that the excess rate of return function is positive and is 
monotonic. In foreign exchange models, numeraire symmetry can be ensured 
by imposing a further inequality on the relation between the volatility and the 
risk aversion. In the extended version of the model, the market 
is driven by a vector of independent L\'evy processes, and the risk 
aversion and volatility coefficients are taken to be predictable processes 
satisfying suitable integrability conditions, these conclusions remain valid. 
Our approach is based on use of the physical measure ${\mathbb P}$. 
We emphasise the importance of the pricing kernel method, because this 
leads to a unified view of the role of L\'evy models in finance, allowing one to 
separate pricing issues from  hedging issues. In particular, we 
make no use of the idea of ``trying to find an equivalent martingale measure" 
by some recipe when one is given a set of price processes. 
Rather,  the pricing kernel is to be regarded as an essential component 
of the theory from the beginning. One needs the pricing kernel before one can 
speak of prices, because the value of a security is determined by the random 
cash flows that it produces, and these need to be valued by use of the pricing 
kernel. 
In this respect our point of view diverges in spirit from the 
earlier literature on L\'evy models in finance, as represented by Gerber \& 
Shiu (1994), Eberlein \& Keller (1995), Eberlein \& Jacod (1997), Chan (1999),  
Raible (2000), Kallsen \& Shiryaev (2002), Fujiwara \& Miyahara (2003), Esche 
\& Schweizer (2005), Eberlein \textit{et al}.~(2005), Hubalek \& Sgarra (2006), 
and others. Exceptions include: Madan \& Milne (1991) who are able to identify the rate of return in their 
study of the VG model; Heston (1993) who with the 
introduction of a ``pricing operator" in his study of the gamma model offers 
a point of view similar in some respects to ours; and Madan (2006) where risks 
are priced by such kernels in a particular equilibrium. 

The approach we have 
outlined for asset pricing in a L\'evy setting with predictable risk aversion and 
volatility is also useful in the theory of interest rates, since it allows one to generalise 
the HJM (Heath \textit{et al}.~1992) framework in a natural way to the L\'evy 
category, without the need of introducing instantaneous forward rates, but in a 
way that guarantees positive excess rates of returns on bonds, and is 
formulated in the  ${\mathbb P}$-measure, making it suitable as a practical 
basis for risk management, forecasting,  and scenario analysis. 

\begin{acknowledgments}
\noindent The authors are grateful to D.~Brigo, M.~Grasselli, S.~Jaimungal, 
A.~Macrina, B.~K.~Meister, B.~Pham-Dang, and  M.~R.~Pistorius for useful discussions. We 
would like to express our gratitude to seminar participants at the Fields Institute, 
Toronto (July 2011), the Japan Association for Financial Engineering and 
Econometrics, Tokyo (August 2011), the Fourth International Conference on 
Mathematics in Finance, Kruger National Park, South Africa (August 2011), 
the Seventh WBS Fixed Income Conference, Berlin (October 2011), and the 
RIO 2011 Research in Options 
Conference, Rio de Janeiro (November 2011) for helpful 
comments. LPH wishes to thank Lloyds TSB Bank Plc, Shell UK Ltd, and the 
Fields Institute, Toronto, for support. EM acknowledges funding by EPSRC.
\end{acknowledgments}




\begin{thebibliography}{999}

\bibitem{applebaum} 
Applebaum,~D. (2004) {\em L\'evy Processes and 
Stochastic Calculus} (Cambridge: Cambridge University Press).
 
\bibitem{bertoin} 
Bertoin,~J. (2004) {\em L\'evy Processes} (Cambridge: Cambridge University Press).

\bibitem{baxter} 
Baxter,~M. (2007) L\'evy simple structural models.~\textit{International 
Journal of Theoretical and Applied Finance} \textbf{10}, 593--606.

\bibitem{billingsley} 
Billingsley,~P. (1995) {\em Probability and Measure} (New Yrok: Wiley).

\bibitem{brody7}
Brody,~D.~C.,~Hughston,~L.~P.~\& Macrina,~A. (2008) 
Dam rain and cumulative gain. {\em Proc.~Roy.~Soc.~Lond.~A} {\bf 464}, 1801--1822.

\bibitem{CGMY} 
Carr,~P., Geman,~H., Madan,~D. \& Yor,~M. (2002) The fine structure of 
asset returns: an empirical investigation. {\em Journal of Business} \textbf{75}, 305--332.

\bibitem{chan}
Chan,~T. (1999) Pricing contingent claims on stocks driven by L\'evy 
processes. {\em Annals of Applied Probability} \textbf{9}, 504--528.

\bibitem{cochrane} 
Cochrane,~J.~H. (2005) {\em Asset Pricing} (Princeton: Princeton University Press).

\bibitem{cont} 
Cont,~R.~\& Tankov,~P.~(2004) {\em Financial Modelling with Jump 
Processes} (London: Chapman \& Hall).

\bibitem{Dicksen} 
Dickson,~D.~C.~M. \& Waters,~H.~R. (1993)
Gamma processes and finite time survival probabilities. {\em ASTIN Bulletin} 
\textbf{23}, 259-272.

\bibitem{Dufresne}
Dufresne,~F., Gerber,~H.~U. \& Shiu,~E.~S. (1991) 
Risk theory with the gamma process. {\em ASTIN Bulletin} \textbf{21}, 177-192.

\bibitem{eberlein1} 
Eberlein,~E. \& Jacod,~J. (1997) On the range of option prices. 
{\em Finance and Stochastics} \textbf{1}, 131--140.

\bibitem{eberlein2} 
Eberlein,~E., Jacod,~J. \& Raible,~S. (2005) L\'evy term structure 
models: No-arbitrage and completeness. 
{\em Finance and Stochastics} \textbf{9}, 67--88.

\bibitem{eberlein3} 
Eberlein,~E. \& Keller,~U. (1995) Hyperbolic distributions in finance. 
{\em Bernoulli} \textbf{1}, 281--299.

\bibitem{eberlein4} 
Eberlein,~E. \& Raible,~S. (1999) Term structure models driven by 
general  L\'evy  processes. {\em Mathematical Finance} \textbf{9}, 31--53.

\bibitem{Esche} 
Esche,~F. \& Schweizer,~M. (2005) Minimal entropy preserves the L\'evy 
property: how and why. {\em Stochastic Processes and their Applications} 
\textbf{115}, 299--327.

\bibitem{Fujiwara} 
Fujiwara,~T. \& Miyahara,~Y. (2003) The minimal entropy martingale 
measures for geometric L\'evy processes. {\em Finance and Stochastics} 
\textbf{7}, 509--531.

\bibitem{Gerber}
Gerber,~H.~U. \& Shiu,~E.~S.~W. (1994) Option pricing by Esscher 
transforms (with discussion). {\em Transactions of the Society of Actuaries} 
\textbf{46}, 99-191.

\bibitem{gordon}
Gordon,~M.~J. (1959) Dividends, earnings and stock prices. {\em Review 
of Economics and Statistics} \textbf{41}, 99--105.

\bibitem{HJM} 
Heath,~D., Jarrow,~R. \& Morton,~A. (1992) Bond pricing and the term 
structure of interest rates: a new methodology for contingent claim valuation. 
{\em Econometrica} \textbf{60}, 77--105.

\bibitem{heston} 
Heston,~S.~L. (1993) Invisible parameters in option prices. 
{\em Journal of Finance} \textbf{48}, 993--947.

\bibitem{Hubalek} 
Hubalek,~F. \& Sgarra,~C. (2006) On the Esscher transform and 
entropy for exponential L\'evy models. {\em Quantitative Finance} \textbf{6}, 
125--145. 

\bibitem{Kallsen}  
Kallsen,~J. \& Shiryaev,~A.~N. (2002) The cumulant process and 
Esscher's change of measure. {\em Finance and Stochastics} \textbf{6}, 97--428.

\bibitem{Kozubowski} 
Kozubowski,~T.~J. \& Krzysztof,~P. (2009)
Distributional properties of the negative binomial L\'evy process. {\em Probability 
and Mathematical Statistics} \textbf{29}, 43--71.

\bibitem{kyprianou} 
Kyprianou,~A.~E. (2006) {\em Introductory Lectures on 
Fluctuations of L\'evy Processes with Applications} (Berlin: Springer). 

\bibitem{madan0} 
Madan,~D.  (2006) Equilibrium asset pricing: with non-Gaussian factors and exponential utilities. {\em Quantitative Finance} \textbf{6}, 
455-463. 

\bibitem{madan}
Madan,~D., Carr,~P., \& Chang,~E.~C. (1998) The variance 
gamma process and option pricing. 
{\em European Finance Review} \textbf{2}, 79--105.

\bibitem{madan1}
Madan,~D. \& Seneta,~E. (1990) The variance gamma (V.G.) 
model for share market returns. {\em Journal of Business} \textbf{63}, 511--524.

\bibitem{madan2} 
Madan,~D. \& Milne,~F. (1991) Option pricing with V.G. martingale 
components. {\em Mathematical Finance} \textbf{1}, 39--55.

\bibitem{merton}
Merton,~R.~C. (1976) Option pricing when underlying stock returns 
are discontinuous. {\em J. Financial Economics} \textbf{3}, 125--144.

\bibitem{protter} 
Protter, P.~(1990) {\em Stochastic Integration and Differential Equations} 
(New York: Springer).

\bibitem{raible} 
Raible,~S. (2000) {\em L\'evy Processes in Finance: Theory, 
Numerics, and Empirical Facts}.  PhD Thesis, University of Freiburg.

\bibitem{sato} 
Sato,~K. (1999) {\em L\'evy Processes and Infinitely Divisible 
Distributions} (Cambridge: Cambridge University Press). 

\bibitem{schoutens}
Schoutens,~W. (2004) {\em L\'evy Processes in Finance: Pricing 
Financial Derivatives} (New York: Wiley).

\bibitem{siegel}
Siegel,~J.~J. (1972) Risk, Interest rates and the forward exchange. 
{\em Quarterly Journal of Economics} \textbf{86}, 303--309.

\bibitem{yor}
Yor,~M. (2007) Some remarkable properties of gamma
processes.~In:~{\em Advances in Mathematical Finance, Festschrift
volume in honour of Dilip Madan}.~R.~Elliott, M.~Fu, R.~Jarrow \&
Ju-Yi~Yen, eds.~(Basel: Birkh\"auser).

\end{thebibliography}
\end{document}